\documentclass[12pt]{article}
\usepackage[centertags]{amsmath}
\usepackage{amsfonts}
\usepackage{amssymb}
\usepackage{graphicx}
\usepackage{amssymb}
\usepackage[verbose,colorlinks=true,naturalnames=true,linkcolor=blue,]
{hyperref}

\newcounter{rown}

\begin{document}

\title{Once again about quantum deformations of $D=4$ Lorentz algebra: twistings of
$q$-deformation}

\author{A. Borowiec$^{1,3}$, J. Lukierski$^{1}$ and V.N. Tolstoy$^{1,2}$
\\ \\
$^{1}$Institute for Theoretical Physics,
\\University of Wroc{\l}aw, pl. Maxa Borna 9,
\\50--205 Wroc{\l}aw, Poland\\
\\$^{2}$Institute of Nuclear Physics,
\\Moscow State University, 119 992 Moscow, Russia\\
\\$^{3}$Bogoliubov Laboratory of Theoretical Physics,
\\Joint Institute for Nuclear Research, Dubna,
\\ Moscow region 141980, Russia}

\date{}

\maketitle
\begin{abstract}
This paper together with the previous one \cite{BLT1} presents the detailed description
of all quantum deformations of $D=4$ Lorentz algebra as Hopf algebra in terms of complex
and real generators. We describe here in detail two quantum deformations of the $D=4$
Lorentz algebra $\mathfrak{o}(3,1)$ obtained by twisting of the standard $q$-deformation
$U_{q}(\mathfrak{o}(3,1))$. For the first twisted $q$-deformation an Abelian twist
depending on Cartan generators of $\mathfrak{o}(3,1)$ is used. The second example of
twisting provides a quantum deformation of Cremmer-Gervais type for the Lorentz algebra.
For completeness we describe also twisting of the Lorentz algebra by standard Jordanian
twist. By twist quantization techniques we obtain for these deformations new explicit
formulae for the deformed coproducts and antipodes of the $\mathfrak{o}(3,1)$-generators.
\end{abstract}

\section{Introduction}

The quantization of gravity is not only important as the completion of the quantum
description of fundamental interactions - it affects also the basic structure of space
time and the nature of relativistic symmetries (see e.g \cite{DFR}-\cite{FrLi}). One can
conjecture that the noncommutative space-time and quantum symmetries, described by
noncommutative Hopf algebra, provide an algebraic deformation of the classical symmetry
framework which is caused by the quantum gravity corrections. In the center of all
relativistic considerations is the Lorentz symmetry, and therefore all possible
modifications of Lorentz symmetries should be carefully studied. The following two ways
of studying the Lorentz symmetry deformations have been proposed:

(i) The simplest way is obtained by considering nonlinear realizations of classical
Lorentz symmetries (see e.g. \cite{MaSm}-\cite{JuVi}) obtained usually by a nonlinear
transformation of the four-momentum basis. In such a way we can mainly interpret the
effects due to the modification of relativistic mass shell condition (see e.g.
\cite{AmCa}). In such a framework the space-time manifold remains commutative, i.e. one
can use the methods of classical geometry and classical group theory.

(ii) The quantum extension of a symmetry group is provided by the noncommutative
quasitriangular Hopf-algebras, with representation spaces described by non-commutative
modules. This technique was extensively applied to relativistic symmetries (see e.g.
\cite{PoWo}-\cite{AschCa}).

In the present paper the second way of modifying relativistic symmetries is investigated.
The formalism of quasitriangular Hopf algebras \cite{Dr1} describing the deformations of
universal enveloping algebras and the corresponding dual quantum groups were extensively
studied (see e.g. \cite{Man}-\cite{ChaPr}) in order to describe quantum  modification of
physical symmetries and introduce the noncommutative geometry in physics. Very important
from the point of view of possible physical applications are the quantum Hopf-algebraic
deformations of the Lorentz and Poincare algebras.

The Hopf-algebraic deformations are described infinitesimally by the Poisson structures,
satisfying homogenous (standard) or inhomogeneous (modified) Yang-Baxter equations. The
Poisson structures for Lorentz algebra are well-known and have been classified some time
ago by S. Zakrzewski \cite{Zak1} (see also \cite{Mu}) who provided basic four classical
$\mathfrak{o}(3,1)$ $r$-matrices. Two $D=4$ Lorentz $r$-matrices generate the Jordanian
and extended Jordanian deformations of $\mathfrak{o}(3,1)$. The extended Jordanian
deformation was considered in detail for complex basis  as well as  real basis of
$\mathfrak{o}(3,1)$ \cite{BLT1,BLT2}. The Jordanian deformation is defined by the
well-known Jordanian twist and Hopf structure of this deformation is presented in this
paper. Remaining two $r$-matrices generate quantum deformations which are less known, and
are studied explicitly in the present paper. We use extensively the property that these
deformations can be described as twisting of $q$-deformed Hopf algebra
$U_q(\mathfrak{o}(3,1))$ \cite{SWZ,KuMu}. It should be stressed that most of twisting
procedures considered in the literature are imposed on the classical Lie algebra
structures. In this paper we consider the twists modifying already quantum-deformed
Hopf-algebraic symmetry\footnote{This inclusion is different from the Hopf-algebraic one
presented by Podle\'{s} and Woronowicz \cite{PoWo}, where $U_q(\mathfrak{sl}(2))$ is
constructed as the extension of $U_q(\mathfrak{su}(2))$ via double product
construction.}.

The plan of our paper is the following. In Sect.2 the complete list of classical
$r$-matrices (see \cite{Zak1}) is presented in terms of real and complex generators.
Subsequently  we describe in Sect.3 the explicit Hopf algebra structure of the Lorentz
algebra quantized by Jordanian twist \cite{Og}. First twisting of the $q$-deformed
Lorentz algebra $U_q(\mathfrak{o}(3,1))$ is obtained by using the Abelian twist in
Sect.4, which is a function of the Cartan subalgebra of $U_q(\mathfrak{o}(3,1))$. In this
case explicit formulae of Hopf structure are given in terms of the complex and real
Cartan-Weyl basis of $U_q(\mathfrak{o}(3,1))$. We show also that the inclusion
$U_q(\mathfrak{o}(3,1))\supseteq U_q(\mathfrak{o}(3))$ can be realized on the algebra
level after suitable choice of the basis for $U_q(\mathfrak{o}(3))$\footnote{For an
analogous twist modifying $\kappa$-deformation of the Poincar\'{e} algebra see
\cite{LuLya}.}. In Sect.5 we use a $q$-Abelian twist and as a result we obtain the
quantum deformation for the Lorentz algebra of Cremmer-Gervais type
\cite{BeDr}-\cite{IsOg}. Explicit formulae of Hopf structure for this case are also given
in terms of the complex and real Cartan-Weyl basis of $U_q(\mathfrak{o}(3,1))$. The
Sect.6 is Outlook. In last Sect.7 we consider some specialization of $q$-Hadamard formula
which describes a similarity transformation by a $q$-exponential.

It should be recalled that the $q$-deformation does not permit the extension of the
deformation  of Lorentz symmetry to $q$-deformed Poincare algebra in the framework of
standard Hopf algebra, with the coproducts defined on standard tensor products. If we
wish to find the deformed counterpart of Poincare algebra extending Drinfeld-Jimbo
deformation of Lorentz algebra, we have to consider the class of braided quantum Poincare
algebras \cite{Maj2}, which  require as well the deformation of tensor categories. On the
other side one can show \cite{Do,LuNo} that the $q$-deformation of the Lorentz symmetry
can be embedded as a Hopf subalgebra in the $q$-deformed Weyl algebra obtained by adding
to the Poincar\'{e} algebra the dilatation generators.

\setcounter{equation}{0}
\section{$D=4$ Lorentz algebra and its classical $r$-matrices}
Firstly we remind some information from \cite{BLT1}. The classical canonical basis of the
$D=4$ Lorentz algebra, $\mathfrak{o}(3,1)$, can be described by anti-Hermitian six
generators ($h$, $e_{\pm}$, $h'$, $e'_{\pm}$) satisfying the following non-vanishing
commutation relations\footnote{Since the real Lie algebra $\mathfrak{o}(3,1)$ is standard
realification of the complex Lie $\mathfrak{sl}(2,\mathbb{C})$ these relations are easy
obtained from the defining relations for $\mathfrak{sl}(2,\mathbb{C})$, i.e. from
(\ref{v1}).}:
\begin{eqnarray}\label{v1}
&[h,\,e_{\pm}^{}]\;=\;\pm e_{\pm}^{}\,,\qquad [e_{+}^{},\,e_{-}^{}]\;=\;2h~,
\\[8pt]\label{v2}
&[h,\,e'_{\pm}]\;=\;\pm e'_{\pm}~,\qquad [h',\,e_{\pm}]\;=\;\pm e'_{\pm}~,\qquad
[e_{\pm}^{},\,e'_{\mp}]\;=\;\pm2h'~ ,
\\[8pt]\label{v3}
& [h',\,e'_{\pm}]\;=\;\mp e_{\pm}^{}~,\qquad [e'_{+},\,e'_{-}]\;=\;-2h~,
\end{eqnarray}
and moreover
\begin{equation}\label{v4}
a^*\;=\;-a\qquad (\forall\;a\;\in\; \mathfrak{o}(3,1))~.\textbf{}
\end{equation}
A complete list of classical $r$-matrices which describe all Poison structures and
generate quantum deformations for $\mathfrak{o}(3,1)$ involve the four independent
formulas \cite{Zak1}:
\begin{eqnarray}\label{v5}
r_{1}^{}\!\!&=\!\!&\alpha\,(e_{+}\wedge h-e'_{+}\wedge h')+2\beta\, e'_{+}\wedge e_{+}~,
\\[5pt]\label{v6}
r_{2}^{}\!\!&=\!\!&\alpha\,e_{+}\wedge h~,
\\[5pt]\label{v7}
r_{3}^{}\!\!&=\!\!&\alpha\,(e'_{+}\wedge e_{-}+ e_{+}\wedge e'_{-})\,+\,
\beta\,(e_{+}\wedge e_{-}\,-\,e'_{+}\wedge e'_{-})-2\gamma\,h\wedge h'~,
\\[5pt]\label{v8}
r_{4}^{}\!\!&=\!\!&\alpha\bigl(e'_{+}\wedge e_{-}+ e_{+}\wedge e'_{-}-2h\wedge
h'\bigr)\pm e_{+}\wedge e'_{+}~.
\end{eqnarray}
All $r$-matrices are skew-symmetric, i.e. $r^{21}_{j}=- r^{12}_{j}$. Moreover if the
universal $R$-matrices $R_{r_{j}^{}}$ ($j=1,2,3,4$) of the quantum deformations
corresponding to the classical $r$-matrices (\ref{v5})--(\ref{v8}) are unitary then these
$r$-matrices are anti-Hermitian, i.e.
\begin{equation}\label{v9}
r^*_{j}\;=\;-r_{j}\qquad (j=1,2,3,4)~.
\end{equation}
Therefore the $*$-operation (\ref{v4}) should be lift to the tensor product
$\mathfrak{o}(3,1)\otimes \mathfrak{o}(3,1)$. There are two variants of this lifting:
{\it direct} and {\it flipped} \cite{To1}, namely,
\begin{eqnarray}\label{v10}
(a\otimes b)^*\!\!&=\!\!&a^*\otimes b^*\qquad({\rm*-direct})~,
\\[5pt]\label{v11}
(a\otimes b)^*\!\!&=\!\!&b^*\otimes a^*\qquad({\rm*-flipped})~.
\end{eqnarray}
We see that if the "direct" lifting of the $*$-operation (\ref{v4}) is used then all
parameters in (\ref{v5})--(\ref{v8}) are pure imaginary. In the case of the "flipped"
lifting (\ref{v11}) all parameters in (\ref{v5})--(\ref{v8}) are real.

The first $r$-matrix (\ref{v5}) satisfies the homogeneous CYBE and it is Jordanian type.
Corresponding quantum deformation for the case (\ref{v10}) was described detailed in the
paper \cite{BLT1} and it is entire defined by the extended Jordanian twit\footnote{This
twists is in accord with the $r$-matrix (\ref{v5}) in the sense of a formula of the type
(\ref{ct18}).}:
\begin{equation}\label{v12}
F_{r_1^{}}^{}\,=\,\exp{\Bigl(\frac{\imath\beta}{\alpha^2}\;\sigma\wedge\varphi\Bigr)}\,
\exp{(h\otimes\sigma-h'\otimes\varphi)}~,
\end{equation}
\begin{equation}\label{v13}
\sigma \,=\,\frac{1}{2}\ln\left[(1+\alpha e_+)^2\!+(\alpha e'_+)^2
\right],\quad\varphi\,=\,\arctan{\frac{\alpha e'_+}{1+\alpha e_+}}
\end{equation}
The second $r$-matrix (\ref{v6}) also satisfies the homogeneous CYBE and it is the
standard Jordanian $r$-matrix. Corresponding twist provided the condition (\ref{v10}) is
presented in the next section.

The last two $r$-matrices (\ref{v7}) and (\ref{v8}) satisfy the non-homogeneous
(modified) CYBE and they can be easy obtained from solutions of the complex algebra
$\mathfrak{o}(4,\mathbb{C})\simeq \mathfrak{sl}
(2,\mathbb{C})\oplus\mathfrak{sl}(2,\mathbb{C})$ which is complexification of
$\mathfrak{o}(3,1)$. Indeed, let us introduce the complex basis of Lorentz algebra
$(\mathfrak{o}(3,1)\simeq\mathfrak{sl}(2;\mathbb{C})\oplus\mathfrak
{\overline{sl}}(2,\mathbb{C}))$ described by two commuting sets of complex generators:
\begin{eqnarray}\label{v14}
H_1\!\!&=\!\!& \frac{1}{2}\,(h+\imath h')~,\qquad
E_{1\pm}\;=\;\frac{1}{{2}}\,(e_{\pm}^{}+ \imath e'_{\pm})~,
\\[5pt]\label{v15}
H_2\!\!&=\!\!& \frac{1}{{2}}\, (h-\imath h')~,\qquad
E_{2\pm}\;=\;\frac{1}{{2}}\,(e_{\pm}^{}-\imath e'_{\pm})~,
\end{eqnarray}
which satisfy the relations (compare with (\ref{v1}))
\begin{equation}\label{v16}
[H_k,\,E_{k\pm}]\;=\;\pm E_{k\pm}~,\qquad [E_{k+},\,E_{k-}]\;=\;2
H_k\qquad(k=1,2)~.
\end{equation}
The $*$-operation describing the real structure  acts on the generators $H_k$, and
$E_{k\pm}$ ($k=1,2$) as follows
\begin{equation}\label{v17}
H_1^*\;=\;-H_2^{}~,\qquad E_{1\pm}^*\;=\;- E_{2\pm}^{}~,\qquad
H_2^*\;=\;-H_1^{}~,\qquad E_{2\pm}^*\;=\;-
E_{1\pm}^{}~.
\end{equation}
The classical $r$-matrix $r_3$ and $r_4$ in terms of the complex basis (\ref{v9}),
(\ref{v10}) take the form
\begin{equation}\label{v18}
\begin{array}{rcl}
r_3^{}\!\!&=\!\!&r'_3+r''_3~,
\\[7pt]
r'_3\!\!&:=\!\!&2(\beta+\imath\alpha)E_{1+}\wedge
E_{1-}+2(\beta-\imath\alpha)E_{2+}\wedge E_{2-}~,
\\[7pt]
r''_3\!\!&:=\!\!&4\imath\gamma\,H_{2}\wedge H_{1}~,
\end{array}
\end{equation}
and
\begin{equation}\label{v19}
\begin{array}{rcl}
r_4^{}\!\!&=\!\!&r'_4+r''_4~,\quad
\\[7pt]
r'_4\!\!&:=\!\!&2\imath\alpha(E_{1+}\wedge E_{1-}-E_{2+}\wedge E_{2-}-2H_{1}\wedge
H_{2})~,
\\[7pt]
r''_4\!\!&:=\!\!&2\imath\nu\,E_{1+}\wedge E_{2+}~.
\end{array}
\end{equation}
For the sake of convenience we introduce parameter $\nu$ in\footnote{We can reduce this
parameter $\nu$ to $\pm1$ by automorphism of $\mathfrak{o}(4,\mathbb{C})$.} $r''_{4}$. It
should be noted that $r'_{3}$, $r''_{3}$ and $r'_{4}$, $r''_{4}$ are themselves classical
$r$-matrices. We see that the $r$-matrix $r'_{3}$ is simply a sum of two standard
$r$-matrices of $\mathfrak{sl}(2;\mathbb{C})$, satisfying the anti-Hermitian condition
$r^*=-r$. Analogously, it is not hard to see that the $r$-matrix $r_4$ corresponds to a
Belavin-Drinfeld triple \cite{BeDr} for the Lie algebra $\mathfrak{sl}(2;\mathbb{C})
\oplus\mathfrak{\overline{sl}}(2,\mathbb{C}))$. Indeed, applying the Cartan automorphism
$E_{2\pm}\rightarrow E_{2\mp}$, $H_{2}\rightarrow-H_{2}$ we see that this is really
correct (see also \cite{IsOg}).

\setcounter{equation}{0}
\section{Hopf structure of Jordanian deformation for Lorentz algebra
$\mathfrak{o}(3,1)$}

Whereas the quantum twist $F_{r_{2}^{}}$ corresponding to the classical Jordanian
$r$-matrix (\ref{v7}) was well known for a long time \cite{Og}
\begin{equation}\label{J1}
F_{r_2^{}}^{}\,=\,\exp{(h\otimes\sigma}),\qquad\sigma\;=\;\ln(1+\alpha e_{+})~,
\end{equation}
however we did not find in a literature any complete Hopf structure for this Jordanian
deformation of the Lorentz algebra $\mathfrak{o}(3,1)$. In this Section we present this
Hopf structure, namely, we give explicit formulas for the co-products
$\Delta_{r_2}(\cdot) :=F_{r_{2}}^{}\Delta(\cdot)F_{r_{2}}^{-1}$ and antipodes
$S_{r_2}(\cdot)=uS(\cdot)u^{-1}$ of the Jordanian deformation of Lorentz algebra
$\mathfrak{o}(3,1)$ for all classical canonical basis ($h$, $e_{\pm}$, $h'$, $e'_{\pm}$).
Here $\Delta(\cdot)$ and $S(\cdot)$ are primitive (non-deformed), i.e.
$\Delta(a)=a\otimes1 +1\otimes a$ and $S(a)=-a$ for $\forall
a\,\in\{h,e_{\pm},h',e_{\pm}\}$), and $u$ is given as follows (see \cite{To2})
\begin{eqnarray}\label{J2}
u\!\!&=\!\!&m({\rm id}\otimes S)(F_{r_{2}^{}})\;=\;\exp{(-\alpha he_+)}~.
\end{eqnarray}
Using a twist technics presented in Section III.D of the paper \cite{BoLuLyaTo} it is not
hard to calculate the  following formulas for the deformed co-products
$\Delta_{r_2}(\cdot)$ for all canonical basis ($h$, $e_{\pm}$, $h'$, $e'_{\pm}$):
\begin{eqnarray}\label{J3}
\Delta _{r_2^{}}(h)\!\!&=\!\!&h\otimes e^{-\sigma }+1\otimes h~,
\\[7pt]\label{J4}
\Delta _{r_2^{}}(h')\!\!&=\!\!&h'\otimes1+1\otimes h'-\alpha h\otimes e'_{+}e^{-\sigma}~,
\\[7pt]\label{J5}
\Delta _{r_2^{}}(e_{+})\!\!&=\!\!&e_{+}\otimes e^{\sigma }+1\otimes e_{+}~,
\\[7pt]\label{J6}
\Delta _{r_2^{}}(e'_{+})\!\!&=\!\!&e'_{+}\otimes e^{\sigma}+1\otimes e'_{+}~,
\\[7pt]\label{J7}
\Delta _{r_2^{}}(e_{-})\!\!&=\!\!&e_{-}\otimes e^{-\sigma}+1\otimes e_{-}+2\alpha
h\otimes he^{-\sigma}-\alpha ^{2}h(h-1)\otimes e_{+}e^{-2\sigma}~,
\\[7pt]\label{J8}
\Delta _{r_2^{}}(e'_{-})\!\!&=\!\!&e'_{-}\otimes e^{-\sigma}+1\otimes e'_{-}+2\alpha
h\otimes h'e^{-\sigma}-\alpha^{2}h(h-1)\otimes e' _{+}e^{-2\sigma}~.
\end{eqnarray}
It should be noted that the formulas (\ref{J3}), (\ref{J5}) and (\ref{J7}) can be found
in \cite{Og} (cf. also \cite{BoLuTo} where superextension of the Jordanian twist has been
described). Using the formula (\ref{J2}) one can easy calculate the formulas of the
deformed antipodes $S_{r_2}(\cdot)$:
\begin{eqnarray}\label{J9}
S_{r_2^{}}(h)\!\!&=\!\!&-he^{\sigma}~,\qquad\quad\; S_{r_2^{}}(h')\;=\;-h'-\alpha
he'_{+}~,
\\[7pt]\label{J10}
S_{r_2^{}}( e_{+})\!\!&=\!\!&- e_{+}e^{-\sigma}~,\qquad
S_{r_2^{}}(e'_{+})\;=\;-e'_{+}e^{-\sigma}~,
\\[7pt]\label{J11}
S_{r_2^{}}(e_{-})\!\!&=\!\!&-e_{-}e^{\sigma}+2\alpha h^{2}e^{\sigma}+
\alpha^{2}h(h-1)e_{+}e^{\sigma}~,
\\[7pt]\label{12}
S_{r_2^{}}(e'_{-})\!\!&=\!\!&-e'_{-}e^{\sigma}+2\alpha hh'e^{\sigma}+\alpha^{2}
h\left(h-1\right)e'_{+}e^{\sigma}~.
\end{eqnarray}

\setcounter{equation}{0}
\section{Twisted $q$-deformation of Cartan type for Lorentz algebra
$\mathfrak{o}(3,1)$ }

In this Section we explicitly describe quantum deformation corresponding to the classical
$r$-matrix $r_{3}$ (\ref{v13}). Since the $r$-matrix $r''_{3}$ is Abelian and it
co-commutes with $r'_{3}$ (see \cite{To1,To2}) therefore we firstly quantize
$\mathfrak{o}(3,1)$ in the direction $r'_{3}$ and then we apply an Abelian twist
corresponding to the $r$-matrix $r''_{3}$.

For the sake of convenience we introduce the following notations
$z_{\pm}:=\beta\pm\imath\alpha$. It should be noted that $z_{-}^{}=z_{+}^*$ if the
parameters $\alpha$ and $\beta$ are real, and  $z_{-}^{}=-z_{+}^*$ if the parameters
$\alpha$ and $\beta$ are pure imaginary. From structure of the classical $r$-matrix
$r_{3}'$, (\ref{v19}), follows that a quantum deformation $U_{r'_{3}}
(\mathfrak{o}(3,1))$ is a combination of two $q$-analogs of
$U(\mathfrak{sl}(2;\mathbb{C}))$ with the parameter $q_{z_{+}^{}}$ and $q_{z_{-}^{}}$,
where $q_{z_{\pm}^{}}:=\exp{z_{\pm}^{}}$. Thus $U_{r'_{3}}(\mathfrak{o}(3,1))\cong
U_{q_{z_{+}^{}}^{}}(\mathfrak{sl}(2;\mathbb{C}))\otimes
U_{q_{z_{-}^{}}^{}}(\mathfrak{sl}(2;\mathbb{C}))$ and the standard generators
$q_{z_{+}^{}}^{\pm H_{1}}$, $E_{1\pm}$ and $q_{z_{-}^{}}^{\pm H_{2}}$, $E_{2\pm}$ satisfy
the following non-vanishing defining relations
\begin{eqnarray}\label{ct1}
q_{z_{+}}^{H_1}E_{1\pm}\!\!&=\!\!&q_{z_{+}}^{\pm1}E_{1\pm}\,q_{z_{+}}^{H_1}~,\qquad
[E_{1+},\,E_{1-}]\;=\;\frac{q_{z_{+}}^{2H_1}-q_{z_{+}}^{-2H_1}}
{q_{z_{+}}^{}-q_{z_{+}}^{-1}}~,
\\[7pt]\label{ct2}
q_{z_{-}}^{H_2}E_{2\pm}\!\!&=\!\!&q_{z_{-}}^{\pm1}E_{2\pm}\,q_{z_{-}}^{H_2}~,\qquad
[E_{2+},\,E_{2-}]\;=\;\frac{q_{z_{-}}^{2H_2}-q_{z_{-}}^{-2H_2}}
{q_{z_{-}}^{}-q_{z_{-}}^{-1}}~.
\end{eqnarray}
In this case the co-product $\Delta_{r'_{3}}$ and antipode $S_{r'_{3}}$ for can be given
by the formulas:
\begin{eqnarray}\label{ct3}
\Delta_{r'_{3}}^{}(q_{z_{+}}^{\pm H_{1}})\!\!&=\!\!&q_{z_{+}}^{\pm H_{1}}\otimes
q_{z_{+}}^{\pm H_{1}}~, \qquad\Delta_{r'_{3}}^{}(E_{1\pm})\;=\;E_{1\pm}\otimes
q_{z_{+}}^{H_{1}}+q_{z_{+}}^{-H_{1}}\otimes E_{1\pm}~,
\\[7pt]\label{ct4}
\Delta_{r'_{3}}^{}(q_{z_{-}}^{\pm H_{2}})\!\!&=\!\!&q_{z_{-}}^{\pm H_{2}}\otimes
q_{z_{-}}^{\pm H_{2}}~,\qquad\Delta_{r'_{3}}^{}(E_{2\pm})\;=\;E_{2\pm}\otimes
q_{z_{-}}^{H_{2}}+q_{z_{-}}^{-H_{2}}\otimes E_{2\pm}~,
\end{eqnarray}
\begin{eqnarray}\label{ct5}
S_{r'_{3}}^{}(q_{z_{+}}^{\pm H_{1}})\!\!&=\!\!&q_{z_{+}}^{\mp H_{1}}~,\qquad
S_{r'_{3}}^{}(E_{1\pm})\;=\;-q_{z_{+}}^{\pm1}E_{1\pm}~,
\\[7pt]\label{ct6}
S_{r'_{3}}^{}(q_{z_{-}^{}}^{\pm H_{2}})\!\!&=\!\!&q_{z_{-}}^{\mp H_{2}}~,\qquad
S_{r'_{3}}^{}(E_{2\pm})\;=\;-q_{z_{-}}^{\,\mp1}E_{2\pm}~.
\end{eqnarray}
The $*$-involution describing the real structure on the generators (\ref{v14}) and
(\ref{v15}) can be adapted to the quantum generators $q_{z}^{\pm H_{1}}$, $E_{1\pm}$ and
$q_{z^*}^{\pm H_{2}}$, $E_{2\pm}$ as follows
\begin{equation}\label{ct7}
(q_{z_{+}}^{\pm H_{1}})^*\;=\;q_{z_{+}^{*}}^{\mp H_{2}},\quad\;E_{1\pm}^*\;=\;-
E_{2\pm}^{}~,\quad\;(q_{z_{-}}^{\pm H_{2}})^*\;=\;q_{z_{-}^{*}}^{\mp H_{1}},
\quad\;E_{2\pm}^*\;=\;- E_{1\pm}^{}~,
\end{equation}
and there exit two $*$-liftings: {\it flip} and {\it direct}, namely,
\begin{eqnarray}\label{ct8}
(a\otimes b)^*\!\!&=\!\!&a^*\otimes b^*\qquad({\rm*-direct})~,
\\[5pt]\label{ct9}
(a\otimes b)^*\!\!&=\!\!&b^*\otimes a^*\qquad({\rm*-flipped})
\end{eqnarray}
for any $a\otimes b\in U_{r'_{3}}(\mathfrak{o}(3,1))\otimes U_{r'_{3}}
(\mathfrak{o}(3,1))$, where $*$-direct involution corresponds to the case of the pure
imaginary parameters $\alpha,\,\beta$ and $*$-flipped involution corresponds to the case
of the real deformation parameters $\alpha,\,\beta$. It should be stressed that the Hopf
structure on $U_{r'_{3}}(\mathfrak{o}(3,1))$, (\ref{ct3})--(\ref{ct6}), satisfy the
consistency conditions under the $*$-involution
\begin{equation}\label{ct10}
\Delta_{r'_{3}}(a^*)\;=\;(\Delta_{r'_{3}}(a))^*,\quad\;\; S_{r'_{3}}((S_{r'_{3}}
(a^*))^{*})\;=\;a\quad(\forall a\in U_{r'_{3}}(\mathfrak{o}(3,1))~.
\end{equation}

The universal $R$-matrix, $R_{r'_{3}}$, which connects the direct
$\Delta_{r'_{3}}^{12}:=\Delta_{r'_{3}}^{}$ and opposite $\Delta_{r'_{3}}^{21}$ coproducts
\begin{eqnarray}\label{ct11}
R_{r'_{3}}\Delta_{r'_{3}}^{12}(a)\!\!&=\!\!&\Delta_{r'_{3}}^{21}(a)R_{r'_{3}}
\quad(\forall a\in U(\mathfrak{o}(3,1))
\end{eqnarray}
has the form:
\begin{eqnarray}\label{ct12}
R_{r'_{3}}\!\!&=\!\!&R_{1}'R_{2}'\,=\,R_{2}'R_{1}'~,
\end{eqnarray}
where
\begin{eqnarray}\label{ct13}
R_{1}'\!\!&=\!\!&\exp_{q_{z_{+}}^{-2}}\Bigl((q_{z_{+}}^{}-q_{z_{+}}^{-1})E_{1+}\,
q_{z_{+}}^{-H_{1}}\otimes q_{z_{+}}^{H_{1}}E_{1-}\Bigr) q_{z_{+}}^{2H_{1}\otimes H_{1}}~,
\end{eqnarray}
and
\begin{eqnarray}\label{ct14}
R_{2}'\!\!&=\!\!&\exp_{q_{z_{-}}^{-2}}\Bigl((q_{z_{-}}^{}-q_{z_{-}}^{-1})
E_{2+}\,q_{z_{-}}^{-H_{2}}\otimes q_{z_{-}}^{H_{2}}E_{2-}\Bigr)
\,q_{z_{-}}^{2H_{2}\otimes H_{2}}
\end{eqnarray}
for  $*$-direct involution, i.e. when the parameters $\alpha$, $\beta$ are pure
imaginary, and
\begin{eqnarray}\label{ct15}
R_{2}'\!\!&=\!\!& \exp_{q_{z_{-}}^{2}}\Bigl((q_{z_{-}}^{-1}-q_{z_{-}}^{})
E_{2-}\,q_{z_{-}}^{-H_{2}}\otimes q_{z_{-}}^{H_{2}}E_{2+}\Bigr)
\,q_{z_{-}}^{-2H_{2}\otimes H_{2}}
\end{eqnarray}
for $*$-flipped involution, i.e. when the parameters $\alpha$, $\beta$ are real. Here we
use the standard definition of $q$-exponential
\begin{eqnarray}\label{ct16}
\exp_{q}(x)\!\!&:=\!\!&\sum_{n\geq0}\,\frac{x^n}{(n)_{q}^{}!}~,
\quad\;(n)_{q}^{}!:=(1)_{q}^{}(2)_{q}^{}\cdots
(n)_{q}^{},\quad(n)_{q}^{}=\frac{1-q^n}{1-q}~.
\end{eqnarray}
The universal $R$-matrix (\ref{ct10}) is unitary
\begin{eqnarray}\label{ct17}
R_{r'_{3}}^{*}\!\!&=\!\!&R_{r'_{3}}^{-1}~.
\end{eqnarray}
In the limit $z\rightarrow0$ we have
\begin{eqnarray}\label{ct18}
R_{r'_{3}}\!\!&=\!\!&1+r_{BD}^{}+O(z^2)~,
\end{eqnarray}
where $r_{BD}^{}$ is the classical Belavin-Drinfeld $r$-matrix
\begin{eqnarray}\label{ct19}
r_{BD}^{}\!\!&=\!\!&2z_{+}\bigl(E_{1+}\otimes E_{1-}+H_1\otimes H_1\bigr)
+2z_{-}\bigl(E_{2+}\otimes E_{2-}+H_2\otimes H_2\bigr)
\end{eqnarray}
for  $*$-direct involution, i.e. when the parameters $\alpha$, $\beta$ are pure
imaginary, and
\begin{eqnarray}\label{ct20}
r_{BD}^{}\!\!&=\!\!&2z_{+}\bigl(E_{1+}\otimes E_{1-}+H_1\otimes H_1\bigr)
-2z_{-}\bigl(E_{2-}\otimes E_{2+}+H_2\otimes H_2\bigr)
\end{eqnarray}
for  $*$-flipped involution, i.e. when the parameters $\alpha$, $\beta$ are real. These
$r$-matrix are not skew-symmetric they satisfy the condition
\begin{eqnarray}\label{ct21}
r_{BD}^{12}+r_{BD}^{21}\!\!&=\!\!&\Omega
\end{eqnarray}
where $\Omega$ is the split anti-Hermitian Casimir element of
$\mathfrak{o}(3,1)$\footnote{Here in (\ref{ct22}) and also in (\ref{ct19}), (\ref{ct20})
the generators $E_{1\pm}$, $E_{2\pm}$, and $e_{\pm}^{}$, $e'_{\pm}$ are not deformed.}
\begin{equation}
\begin{array}{rcl}\label{ct22}
\Omega\!\!&=\!\!&2z_{+}\bigl(E_{1+}\otimes E_{1-}+E_{1-}\otimes E_{1+}+2H_1\otimes
H_1\bigr)
\\[7pt]
&&\!\!-2z_{+}^*\bigl(E_{2+}\otimes E_{2-}+E_{2-}\otimes E_{2+}+2H_2\otimes H_2\bigr)
\\[10pt]
\!\!&=\!\!&\imath(z_{+}\!+z_{+}^*)(e_{+}\otimes e'_{-}\!+ e'_{-}\otimes e_{+}\!+
e'_{+}\otimes e_{-}\!+ e_{-}\otimes e'_{+}\!+h\otimes h'\!+h'\otimes h)
\\[7pt]
&&\!\!+(z_{+}\!-z_{+}^*)(e_{+}\otimes e_{-}\!+e_{-}\otimes e_{+}\!-e'_{+}\otimes
e'_{-}\!-e'_{-}\otimes e'_{+}\!+h\otimes h\!-h'\otimes h')~.
\end{array}
\end{equation}
The Belavin-Drinfeld $r$-matrix $r_{BD}^{}$ satisfies the homogeneous classical
Yang-Baxter equation and the $r$-matrix $r_3'$ is a skew-symmetric part of $r_{BD}^{}$,
namely
\begin{eqnarray}\label{ct23}
r_{BD}^{}\!\!&=\!\!& \frac{1}{2}\,r'_3+\frac{1}{2}\,\Omega~.
\end{eqnarray}

Now we consider deformation of the quantum algebra $U_{r'_{3}}(\mathfrak{o}(3,1))$
(secondary quantization of $U(\mathfrak{o}(3,1))$) corresponding to the additional
$r$-matrix $r''_{3}$, (\ref{v18}). Since the generators $H_{1}$ and $H_{2}$ have the
trivial coproduct
\begin{eqnarray}\label{ct24}
\Delta_{r'_{3}}(H_{k})\!\!&=\!\!&H_{k}\otimes 1+1\otimes H_{k}\quad(k=1,2)~,
\end{eqnarray}
therefore the unitary two-tensor
\begin{eqnarray}\label{ct25}
F_{r_{3}''}\!\!:=\!\!&q_{\imath\gamma}^{H_{1}\wedge H_{2}}\qquad
(F_{r_{3}''}^*\;=\;F_{r_{3}''}^{-1})
\end{eqnarray}
satisfies the cocycle condition (see \cite{Dr2})
\begin{equation}\label{ct26}
F^{12}(\Delta_{r'_{3}}\otimes{\rm id})(F)\;=\;F^{23}({\rm id}\otimes\Delta_{r'_{3}})(F)~,
\end{equation}
and the "unital" normalization condition
\begin{equation}\label{ct27}
(\epsilon \otimes{\rm id})(F)\;=\;({\rm id}\otimes\epsilon)(F)=1~,
\end{equation}
where $\epsilon$ is a counit. Thus the complete deformation corresponding to the
$r$-matrix $r_{3}^{}$ is the twisted deformation of $U_{r'_{3}}(\mathfrak{o} (3,1))$,
i.e. the resulting coproduct $\Delta_{r_{3}}^{}$ is given as follows
\begin{eqnarray}\label{ct28}
\Delta_{r_{3}^{}}^{}(a)\!\!&=\!\!&F_{r_{3}''}^{}\Delta_{r_{3}'}^{}(a)
F_{r_{3}''}^{-1}\quad(\forall a\in U_{r'_{3}}(\mathfrak{o}(3,1))~,
\end{eqnarray}
and in this case the resulting antipode $S_{r_{3}^{}}^{}$ does not change,
$S_{r_{3}^{}}^{}=S_{r'_{3}}^{}$. Applying the twisting two-tensor (\ref{ct25}) to the
formulas (\ref{ct3}) and (\ref{ct4}) we obtain
\begin{eqnarray}\label{ct29}
\Delta_{r_{3}^{}}(q_{z_{+}}^{\pm H_{1}})\!\!&=\!\!&q_{z_{+}}^{\pm H_{1}}\otimes
q_{z_{+}}^{\pm H_{1}},\quad\Delta_{r_{3}}(q_{z_{-}}^{\pm H_{2}})\;=\;q_{z_{-}}^{\pm
H_{2}}\otimes q_{z_{-}}^{\pm H_{2}},
\\[7pt]\label{ct30}
\Delta_{r_{3}^{}}(E_{1\pm})\!\!&=\!\!&E_{1\pm}\otimes q_{z_{+}}^{H_{1}}
q_{\imath\gamma}^{\pm H_{2}}+q_{z_{+}}^{-H_{1}}q_{\imath\gamma}^{\mp H_{2}}\otimes
E_{1\pm}~,
\\[7pt]\label{ct31}
\Delta_{r_{3}}(E_{2\pm})\!\!&=\!\!&E_{2\pm}\otimes q_{z_{-}}^{H_{2}}
q_{\imath\gamma}^{\mp H_{1}}+q_{z_{-}}^{-H_{2}}q_{\imath\gamma}^{\pm H_{1}}\otimes
E_{2\pm}~.
\end{eqnarray}
The universal $R$-matrix, $R_{r_{3}^{}}$, corresponding to the resulting $r$-matrix
$r_{3}^{}$, has the form
\begin{eqnarray}\label{ct32}
R_{r_{3}^{}}\!\!&=\!\!&q_{\imath\gamma}^{H_{2}\wedge H_{1}}R_{r'_{1}}
q_{\imath\gamma}^{H_{2}\wedge H_{1}}\;=\;R_{1}^{}R_{2}^{} q_{\imath\gamma}^{2H_{2}\wedge
H_{1}}\;= \;R_{2}^{}R_{1}^{} q_{\imath\gamma}^{2H_{2}\wedge H_{1}}~,
\end{eqnarray}
where
\begin{eqnarray}\label{ct33}
R_{1}^{}\!\!&=\!\!&\exp_{q_{z_{+}}^{-2}}\Bigl((q_{z_{+}}^{}-q_{z_{+}}^{-1})E_{1+}
q_{z_{+}}^{-H_{1}} q_{\imath\gamma}^{-H_{2}}\otimes q_{z_{+}}^{H_{1}}
q_{\imath\gamma}^{-H_{2}} E_{1-}\Bigr)\,q_{z_{+}}^{2H_{1}\otimes H_{1}}~,
\end{eqnarray}
and
\begin{eqnarray}\label{ct34}
R_{2}\!\!&=\!\!&\exp_{q_{z_{-}}^{-2}}\Bigl((q_{z_{-}}-q_{z_{-}}^{-1})
E_{2+}q_{z_{-}}^{-H_{2}}q_{\imath\gamma}^{H_{1}}\otimes q_{z_{-}}^{-H_{2}}
q_{\imath\gamma}^{H_{1}}E_{2-}\Bigr)\,q_{z_{-}}^{2H_{2}\otimes H_{2}}
\end{eqnarray}
for  $*$-direct involution, i.e. when the parameters $\alpha$, $\beta$ are pure
imaginary, and
\begin{eqnarray}\label{ct35}
R_{2}\!\!&=\!\!&\exp_{q_{z_{-}}^{2}}\Bigl((q_{z_{-}}^{-1}-q_{z_{-}})
E_{2-}q_{z_{-}}^{H_{2}}q_{\imath\gamma}^{H_{1}}\otimes q_{z_{-}}^{-H_{2}}
q_{\imath\gamma}^{H_{1}}E_{2+}\Bigr)\,q_{z_{-}}^{-2H_{2}\otimes H_{2}}
\end{eqnarray}
for $*$-flipped involution, i.e. when the parameters $\alpha$, $\beta$ are real. It is
evident that the universal $R$-matrix (\ref{ct32}) is also unitary
\begin{eqnarray}\label{ct36}
R_{r_{3}^{}}^{*}\!\!&=\!\!&R_{r_{3}^{}}^{-1}~.
\end{eqnarray}
In the limit $z\rightarrow0,\,\gamma\rightarrow0$ we have (cf. (\ref{ct12}),
(\ref{ct18}))
\begin{eqnarray}\label{ct37}
R_{r_{3}^{}}\!\!&=\!\!&1+\tilde{r}_{BD}^{}+O(z^2,z\gamma,\gamma^2)~,
\end{eqnarray}
where
\begin{eqnarray}\label{ct38}
\tilde{r}_{BD}^{}\!\!&=\!\!& \frac{1}{2}\,r_3^{}+\frac{1}{2}\,\Omega~.
\end{eqnarray}

Now we introduce a deformed analog of the real canonical basis in the quantum algebra
$U_{r'_{3}}(\mathfrak{o}(3,1))$ by formulas similar to the non-deformed case (\ref{v14})
and (\ref{v15}), namely
\begin{eqnarray}\label{ct39}
h\!\!&=\!\!&H_1+H_2~,\qquad\;\;e_{\pm}\;=\;\sqrt[4]{\lambda{\lambda^*}^{-1}}\,
E_{1\pm}^{}+\Bigl(\sqrt[4]{\lambda{\lambda^*}^{-1}}\Bigr)^*\,E_{2\pm}~,
\\[5pt]\label{ct40}
h'\!\!&=\!\!&\imath(H_2-H_1)~,\quad\;\;e'_{\pm}\;=\;\imath\Bigl(\sqrt[4]
{\lambda{\lambda^*}^{-1}}\Bigr)^*\,E_{2\pm}^{}-\imath\sqrt[4]{\lambda{\lambda^*}^{-1}}\,
E_{1\pm}~.
\end{eqnarray}
where entering the root factor $\sqrt[4]{\lambda{\lambda^*}^{-1}}$ is a matter of
convenience, and $\lambda:=q_{z_{+}}^{}-q_{z_{+}}^{-1}$. These basis elements are
anti-Hermitian, $a^*=-a$ ($a\in\{h,h',e_{\pm}, e'_{\pm}\}$). It is evident that the
commutation relations between the elements $h,\;h'$ and $e_{\pm}^{},\;e_{\pm}^{}$ are not
deformed, that is
\begin{eqnarray}\label{ct41}
[h,\,e_{\pm}^{}]\!\!&=\!\!&[e_{\pm}',\,h']\;=\;e_{\pm}^{}~,\qquad
[h,\,e'_{\pm}]\;=\;[h',\,e_{\pm}]\;=\;e'_{\pm}~.
\end{eqnarray}
Using (\ref{ct1}) and (\ref{ct2}) we obtain the following commutation relations between
the elements $e_{+}^{},\;e'_{+}$, $e_{-}^{},\;e'_{-}$:
\begin{eqnarray}\label{ct42}
[e_{+}^{},\,e_{+}']\!\!&=\!\!&[e_{-}^{},\,e_{-}']\,=\,0~,
\end{eqnarray}
\begin{eqnarray}\label{ct43}
[e_{+}^{},\,e_{-}^{}]\!\!&=\!\!&[e_{-}',\,e_{+}']\;=\;\frac{2\sinh(xh+yh')\cosh(\imath
yh+\imath xh')}{\sqrt{\cosh2x-\cosh2\imath y}}~,
\\[5pt]\label{ct44}
[e_{+}',\,e_{-}^{}]\!\!&=\!\!&[e_{+},\,e'_{-}]\;=\;\frac{-2\imath\sinh(\imath yh+\imath
xh')\cosh(xh+yh')}{\sqrt{\cosh2x-\cosh2\imath y}}~,
\end{eqnarray}
where $x=\mathop{Re}z_{+}$, $y=\mathop{Im}z_{+}$, i.e. $x=\beta$, $y=\alpha$ if $\beta$,
$\alpha$ are real and $x=-\imath\alpha$, $y=-\imath\beta$ if $\beta$, $\alpha$ are pure
imaginary. It should be noted that the morphism $\omega(h)=-h$, $\omega(h')=-h'$,
$\omega(e_{\pm})=-e_{\mp}$, $\omega(e_{\pm}')=-e_{\mp}'$ is automorphism, i.e. $\omega$
is the Cartan automorphism. Using the expressions (\ref{ct39}), (\ref{ct40}),
(\ref{ct29})--(\ref{ct31}) and (\ref{ct3}), (\ref{ct4}) we obtain the following formulas
of the coproducts:
\begin{eqnarray}\label{ct45}
\Delta_{r_{3}^{}}(h)\!\!&=\!\!&h\otimes1+1\otimes h~,\qquad\qquad
\Delta_{r_{3}^{}}(h')\;=\;h'\otimes1+1\otimes h'~,\
\end{eqnarray}
\begin{eqnarray}\label{ct46}
\begin{array}{rcl}
\Delta_{r_{3}^{}}(e_{\pm})\!\!&=\!\!&\displaystyle\frac{1}{2}\Bigl(e_{\pm}\otimes
\bigl(q_{z_{+}}^{H_{1}} q_{\imath\gamma}^{\pm H_{2}}+q_{z_{-}}^{H_{2}}
q_{\imath\gamma}^{\mp H_{1}}\bigr)+\bigl(q_{z_{+}}^{-H_{1}} q_{\imath\gamma}^{\mp H_{2}}+
q_{z_{-}}^{-H_{2}} q_{\imath\gamma}^{\pm H_{1}}\bigr)\otimes e_{\pm}\Bigr)\;+
\\[10pt]
&&+\;\displaystyle\frac{\imath}{2}\Bigl(e_{\pm}'\otimes\bigl(q_{z_{+}}^{H_{1}}
q_{\imath\gamma}^{\pm H_{2}}-q_{z_{-}}^{H_{2}} q_{\imath\gamma}^{\mp H_{1}}\bigr)+
\bigl(q_{z_{+}}^{-H_{1}}q_{\imath\gamma}^{\mp H_{2}}- q_{z_{-}}^{-H_{2}}
q_{\imath\gamma}^{\pm H_{1}}\bigr)\otimes e_{\pm}'\Bigr)~,
\end{array}
\end{eqnarray}
\begin{eqnarray}\label{ct47}
\begin{array}{rcl}
\Delta_{r_{3}^{}}(e_{\pm}')\!\!&=\!\!&\displaystyle\frac{1}{2}\Bigl(e_{\pm}'\otimes
\bigl(q_{z_{+}}^{H_{1}}q_{\imath\gamma}^{\pm H_{2}}+q_{z_{-}}^{H_{2}}
q_{\imath\gamma}^{\mp H_{1}}\bigr)+ \bigl(q_{z_{+}}^{-H_{1}}q_{\imath\gamma}^{\mp H_{2}}+
q_{z_{-}}^{-H_{2}} q_{\imath\gamma}^{\pm H_{1}}\bigr)\otimes e_{\pm}'\Bigr)\;+
\\[10pt]
&&-\;\displaystyle\frac{\imath}{2}\Bigl(e_{\pm}\otimes\bigl(q_{z_{+}}^{H_{1}}
q_{\imath\gamma}^{\pm H_{2}}-q_{z_{-}}^{H_{2}} q_{\imath\gamma}^{\mp H_{1}}\bigr)+
\bigl(q_{z_{+}}^{-H_{1}} q_{\imath\gamma}^{\mp H_{2}}-q_{z_{-}}^{-H_{2}}
q_{\imath\gamma}^{\pm H_{1}}\bigr)\otimes e_{\pm}\Bigr)~,
\end{array}
\end{eqnarray}
where $H_1\,=\,\frac{1}{2}(h+\imath h')$, $H_2\,=\,\frac{1}{2}(h-\imath h')$. The
antipodes are given as follows
\begin{eqnarray}\label{ct48}
S_{r_{3}}(h)\!\!&=\!\!&-h~,\qquad
S_{r_{3}}(e_{\pm}^{})\;=\;-\frac{1}{2}\bigl(q_{z_+}^{\pm1}+q_{z_+^*}^{\pm1}\bigr)
e_{\pm}^{}\mp\frac{\imath}{2}\bigl(q_{z_+}^{\pm1}-q_{z_+^*}^{\pm1}\bigr)e_{\pm}'~,
\\[7pt]\label{ct49}
S_{r_{3}}(h')\!\!&=\!\!&-h'~,\qquad
S_{r_{4}}(e_{+}')\;=\;-\frac{1}{2}\bigl(q_{z_+}^{\pm1}+q_{z_+^*}^{\pm1}\bigr)
e_{\pm}'\pm\frac{\imath}{2}\bigl(q_{z_+}^{\pm1}-q_{z_+^*}^{\pm1}\bigr)e_{\pm}~.
\end{eqnarray}
Turning back to the relation (\ref{ct43}) it is natural to ask whether there exist
another real basis vectors $\tilde{e}_{\pm}$ $\tilde{e}_{\pm}'$ in which the right side
of (\ref{ct43}) would be a function of only Cartan element $h$. The answer is positive if
and only if $y=0$. This basis is given as follows
\begin{eqnarray}\label{ct50}
\tilde{h}\!\!&=\!\!&h~,\qquad\;\tilde{e}_{\pm}^{}\;=\;\frac{1}{2}(e_{\pm}^{}+\imath
e_{\pm}')q_{x}^{H_{2}}+\frac{1}{2}(e_{\pm}^{}-\imath e_{\pm}')q_{x}^{-H_{1}}~,
\\[5pt]\label{ct51}
\tilde{h}'\!\!&=\!\!&h'~,\qquad\tilde{e}_{\pm}'\;=\;\frac{1}{2}(e_{\pm}'+\imath
e_{\pm})q_{x}^{-H_{1}}+\frac{1}{2}(e_{\pm}'-\imath e_{\pm}^{})q_{x}^{H_{2}}~,
\end{eqnarray}
and it has the commutation relations
\begin{eqnarray}\label{ct52}
[\tilde{h},\,\tilde{e}_{\pm}^{}]\!\!&=\!\!&[\tilde{e}_{\pm}',\,\tilde{h}']\;=\;
\pm\;\tilde{e}_{\pm}^{}~,\qquad\qquad\, [\tilde{h},\,\tilde{e}'_{\pm}]\;=\;
[\tilde{h}',\,\tilde{e}_{\pm}]\;=\;\pm\;\tilde{e}'_{\pm}~,
\\[7pt]\label{ct53}
[\tilde{e}_{+}^{},\,\tilde{e}_{-}^{}]\!\!&=\!\!&[\tilde{e}_{-}',\,\tilde{e}_{+}']\;=\;
\frac{\sinh2x\tilde{h}}{\sinh x}~,\qquad[\tilde{e}_{\pm}^{},\,\tilde{e}_{\pm}']\;=
\;\pm\imath\tanh\frac{x}{2}\;(\tilde{e}_{\pm}^{2}+ \tilde{e}_{\pm}^{\,'2})~,
\\[7pt]\label{ct54}
[\tilde{e}_{+}',\,\tilde{e}_{-}^{}]\!\!&=\!\!&[\tilde{e}_{+},\,\tilde{e}'_{-}]\;=\;
\imath\,\frac{\exp(-2\imath x\tilde{h}')- \cosh2x\tilde{h}}{\sinh x}~.
\end{eqnarray}
We see that the commutation relations of the elements $\tilde{h}$ and $\tilde{e}_{\pm}$
are the quantum $q$-ana\-log of the relations (\ref{v1}), therefore the quantum algebra
$U_{q}(\mathfrak{so}(3))$ generated by these elements is the subalgebra of
$U_{q}(\mathfrak{o}(3,1))$,  $U_{q}(\mathfrak{so}(3))\subset U_{q}(\mathfrak{o}(3,1))$,
for a real deformation parameter $q$, $q\in \mathbb{R}$. It should be noted that the
morphism $\omega(\tilde{h})=-h$, $\omega(\tilde{h}')=-h'$,
$\omega(\tilde{e}_{\pm})=-\tilde{e}_{\mp}$, $\omega(\tilde{e}_{\pm}')=-\tilde{e}_{\mp}'$
is not automorphism because it does not  retain the relation (\ref{ct54}). Using the
formulas (\ref{ct52}), (\ref{ct53}) and (\ref{ct46}), (\ref{ct47}) is not difficult to
obtained formulas for the coproduct of the tilled generators. We will not write down
these formulas, however, it is important to note that the subalgebra
$U_{q}(\mathfrak{so}(3))$ is not any Hopf subalgebra of $U_{q}(\mathfrak{o}(3,1))$, i.e.
$\Delta_{r_{3}^{}}(U_{q} (\mathfrak{so}(3)))$ does not belong to $U_{q}(\mathfrak{so}(3))
\otimes U_{q}(\mathfrak{so}(3))$.


\setcounter{equation}{0}
\section{Twisted $q$-deformation of Lorentz algebra $\mathfrak{o}(3,1)$,
corresponding to Belavin-Drinfeld triple}

Next, we describe quantum deformation corresponding to the classical $r$-matrix $r_{4}$
(\ref{v19}). Since the $r$-matrix $r_{4}'(\alpha):=r_{4}'$ is a particular case of
$r_{3}^{}(\alpha,\beta,\gamma):=r_{3}^{}$, namely
$r_{4}'(\alpha)=r_{3}^{}(\alpha,\beta=0,\gamma=\alpha)$, therefore a quantum deformation
corresponding to the $r$-matrix $r_4'$ is obtained from the previous case by setting
$\beta=0,\gamma=\alpha$. Thus, the quantum deformation $U_{r'_{4}}(\mathfrak{o}(3,1))$ is
generated by the elements $q_{\xi}^{\pm H_{1}}$, $E_{1\pm}$ and $q_{\xi}^{\pm H_{2}}$,
$E_{2\pm}$ with the following defining relations
\begin{eqnarray}\label{BD1}
q_{\xi}^{H_k}E_{k\pm}\!\!&=\!\!&q_{\xi}^{\pm1}E_{k\pm}\,q_{\xi}^{H_k}~,\qquad
[E_{k+},\,E_{k-}]\;=\;\frac{q_{\xi}^{2H_k}-q_{\xi}^{-2H_k}}
{q_{\xi}^{}-q_{\xi}^{-1}}\quad\;\; (k=1,2)~.
\end{eqnarray}
Here and elsewhere we will set $\xi:=\imath\alpha$, $\eta:=-2\imath\nu$ and the
parameters $\xi$ and $\eta$ can be simultaneously either real or pure imaginary. The
co-product $\Delta_{r'_{4}}$ and antipode $S_{r'_{4}}$ are given by the formulas:
\begin{eqnarray}\label{BD2}
\Delta_{r_{4}'}(q_{\xi}^{\pm H_{k}})\!\!&=\!\!&q_{\xi}^{\pm H_{k}}\otimes q_{\xi}^{\pm
H_{k}} \qquad (k=1,2)~,
\\[7pt]\label{BD3}
\Delta_{r_{4}'}(E_{1\pm})\!\!&=\!\!&E_{1\pm}\otimes q_{\xi}^{H_{1}\pm H_{2}}+
q_{\xi}^{-H_{1}\mp H_{2}}\otimes E_{1\pm}~,
\\[7pt]\label{BD4}
\Delta_{r_{4}'}(E_{2\pm})\!\!&=\!\!&E_{2\pm}\otimes q_{\xi}^{\mp H_{1}-H_{2}}+
q_{\xi}^{\pm H_{1}+H_{2}}\otimes E_{2\pm}~,
\\[7pt]\label{BD5}
S_{r'_{4}}^{}(q_{\xi}^{\pm H_{k}})\!\!&=\!\!&q_{\xi}^{\mp H_{k}}\qquad (k=1,2)~,
\\[7pt]\label{BD6}
S_{r'_{4}}^{}(E_{1\pm})\!\!&=\!\!&-q_{\xi}^{\pm1}E_{1\pm}~,\quad
S_{r'_{4}}^{}(E_{2\pm})\;=\;-q_{\xi}^{\,\mp1}E_{2\pm}~.
\end{eqnarray}

Now we want to construct deformation of the quantum algebra
$U_{r'_{4}}(\mathfrak{o}(3,1))$ (secondary quantization of $U(\mathfrak{o}(3,1))$)
corresponding to the additional $r$-matrix $r''_{4}$, (\ref{v19}). We describe here only
the pure imaginary case when the parameters $\xi$ and $\eta$ are pure imaginary, i.e.
this case corresponds to the $*$-flipped involution (\ref{ct9})\footnote{The real case
corresponding to the $*$-direct involution (\ref{ct8}) (when $\xi$ and $\eta$ are real)
is rather complicated and it will be described in another place.}. Consider the
two-tensor
\begin{eqnarray}\label{BD7}
F_{r_{4}''}\!\!&:=\!\!&\exp_{q_{\xi}^{2}}^{}\big(\eta E_{1+}q_{\xi}^{H_{1}+H_{2}} \otimes
q_{\xi}^{H_{1}+H_{2}}E_{2+}\big)~.
\end{eqnarray}
It is easy to see that this two-tensor is unitary, $F_{r_{4}''}^*=F_{r_{4}''}^{-1}$, with
respect to the $*$-flipped involution. Moreover, using properties of $q$-exponentials
(see \cite{KhTo1}) is not hard to verify that $F_{r_{4}''}$ satisfies the cocycle
equation (\ref{ct26}). Thus the quantization corresponding to the $r$-matrix $r_4$ is the
twisted $q$-deformation $U_{r_{4}'}(\mathfrak{o}(3,1))$.

Explicit formulas of the co-products $\Delta_{r_{4}^{}}^{}(\cdot)= F_{r_{4}''}^{}
\Delta_{r_{4}'}^{}(\cdot)F_{r_{4}''}^{-1}$ in the complex Cartan-Weyl bases of
$U_{r_{4}'}(\mathfrak{o}(3,1))$ are given as follows (see Appendix)
\begin{eqnarray}\label{BD8}
\Delta_{r_{4}}(q_{\xi}^{\pm(H_{1}-H_2)})\!\!&=\!\!&q_{\xi}^{\pm(H_{1}-H_2)}\otimes
q_{\xi}^{\pm(H_{1}-H_2)}~,
\\[5pt]\label{BD9}
\Delta_{r_{4}}(\,q_{\xi}^{\;\,H_{1}+H_2})\!\!&=\!\!&\mathbb{X}^{-1}\,
q_{\xi}^{H_{1}+H_2}\otimes q_{\xi}^{H_{1}+H_2}~,
\\[7pt]\label{BD10}
\Delta_{r_{4}}(q_{\xi}^{\;-H_{1}-H_2})\!\!&=\!\!&q_{\xi}^{\,-H_{1}-H_2}\otimes
q_{\xi}^{-H_{1}-H_2}\,\mathbb{X}~,
\\[7pt]\label{BD11}
\Delta_{r_{4}}(E_{1+})\!\!&=\!\!&E_{1+}\otimes q_{\xi}^{H_{1}+H_{2}}+
q_{\xi}^{-H_{1}-H_{2}} \otimes E_{1+}\,\mathbb{X}~,
\\[7pt]\label{BD12}
\Delta_{r_{4}}(E_{2+})\!\!&=\!\!&E_{2+}\otimes q_{\xi}^{-H_{1}-H_{2}}\mathbb{X}+
q_{\xi}^{H_{1}+H_{2}}\otimes E_{2+}~,
\end{eqnarray}
\begin{eqnarray}\label{BD13}
\begin{array}{rcl}
\Delta_{r_{4}}(E_{1-})\!\!&=\!\!&E_{1-}\otimes q_{\xi}^{H_{1}-H_{2}}+
q_{\xi}^{H_{2}-H_{1}}\otimes E_{1-}\,-
\\[12pt]
&&-\;\displaystyle\frac{\eta}{q_{\xi}^{}-q_{\xi}^{-1}}\,\bigl(q_{\xi}^{-4H_{1}}\otimes1-
\mathbb{X}^{-1}\bigr)\bigl(q_{\xi}^{3H_{1}+H_{2}}\otimes E_{2+}q_{\xi}^{2H_1}\bigr)~,
\end{array}
\end{eqnarray}
\begin{eqnarray}\label{BD14}
\begin{array}{rcl}
\Delta_{r_{4}}(E_{2-})\!\!&=\!\!&E_{2-}\otimes q_{\xi}^{H_{1}-H_{2}}+
q_{\xi}^{H_{2}-H_{1}}\otimes E_{2-}\,-
\\[12pt]
&&-\;\displaystyle\frac{\eta}{q_{\xi}^{}-q_{\xi}^{-1}}\,\bigl(1\otimes q_{\xi}^{-4H_{2}}-
\mathbb{X}^{-1}\bigr)\bigr(E_{1+}q_{\xi}^{2H_2}\otimes q_{\xi}^{H_{1}+3H_{2}}\bigr)~,
\end{array}
\end{eqnarray}
where
\begin{eqnarray}\label{BD15}
\mathbb{X}\!\!&:=\!\!&1+\eta(q_{\xi}^{2}-1)E_{1+}q_{\xi}^{H_{1}+H_{2}}\otimes
q_{\xi}^{H_{1}+H_{2}} E_{2+}~.
\end{eqnarray}
Explicit formulas of antipodes $S_{r_4}(\cdot)=uS_{r_{4}'}(\cdot)u^{-1}$ where
\begin{eqnarray}\label{BD16}
u^{-1}\!\!&=\!\!&m\circ (S_{r_{4}'}\otimes\mathop{\rm
id})\exp_{q_{\xi}^{2}}^{}\big(\eta\,E_{1+}q_{\xi}^{H_{1}+H_{2}}\otimes
q_{\xi}^{H_{1}+H_{2}}E_{2+}\big)\,=\,\exp_{q_{\xi}^{2}}^{}\big(\!\eta\,
E_{1+}E_{2+}\big)~,
\end{eqnarray}
are given as follows
\begin{eqnarray}\label{BD17}
S_{r_{4}}(q_{\xi}^{H_{1}-H_2})\!\!&=\!\!&q_{\xi}^{-(H_{1}-H_2)}~,\qquad
S_{r_{4}}(q_{\xi}^{-(H_{1}-H_2)})\;=\;q_{\xi}^{H_{1}-H_2}~,
\\[7pt]\label{BD18}
S_{r_{4}}(\,q_{\xi}^{H_{1}+H_2})\!\!&=\!\!&q_{\xi}^{-H_{1}-H_2}X^{-1}~,\quad
S_{r_{4}}(q_{\xi}^{\;-H_{1}-H_2})\;=\;X\,q_{\xi}^{H_{1}+H_2}~,
\\[7pt]\label{BD19}
S_{r_{4}}(E_{1+})\!\!&=\!\!&-q_{\xi}^{}E_{1+}~,\qquad\qquad\quad\;
S_{r_{4}}(E_{2+})\;=\;-q_{\xi}^{-1}E_{2+}~,
\\[7pt]\label{BD20}
S_{r_{4}}(E_{1-})\!\!&=\!\!&-q_{\xi}^{-1}E_{1-}+\frac{\eta}{q_{\xi}^{2}-1}\,
E_{2+}q_{\xi}^{-2H_1}\bigl(q_{\xi}^{4H_1}-X^{-1}\bigr)~,
\\[7pt]\label{BD21}
S_{r_{4}}(E_{2-})\!\!&=\!\!&-q_{\xi}^{}E_{2-}-\frac{\eta}{q_{\xi}^{-2}-1}\,E_{1+}q_{\xi}^{-2H_2}
\bigl(q_{\xi}^{4H_2}-X^{-1}\bigr)~,
\end{eqnarray}
where
\begin{eqnarray}\label{BD22}
X\!\!&:=\!\!&1+\eta(q_{\xi}^{2}-1)E_{1+}E_{2+}~.
\end{eqnarray}

Using the expressions (\ref{ct39}), (\ref{ct40}), and (\ref{BD8})--(\ref{BD15}) we obtain
the following formulas of the coproducts in terms of the real canonical basis:
\begin{eqnarray}\label{BD23}
\Delta_{r_{4}}(q_{\alpha}^{\pm h'})\!\!&=\!\!&q_{\alpha}^{\pm h'}\otimes q_{\alpha}^{\pm
h'}~,{\hskip 9.5cm} \phantom{aa}
\end{eqnarray}
\begin{eqnarray}\label{BD24}
\begin{array}{rcl}
\Delta_{r_{4}}(q_{\imath\alpha}^{\mp h})\!\!&=\!\!&\Bigl(q_{\imath\alpha}^{-h}\otimes
q_{\imath\alpha}^{-h}\,+
\\[8pt]
&&+\,\nu\sin\alpha\;\bigl(e_{+}\otimes e_{+}\!+e_{+}'\otimes e_{+}'\!+\imath
e_{+}'\otimes e_{+}\!-\imath e_{+}\otimes e_{+}'\bigr)\Bigr)^{\pm1}\!\!,\phantom{aaaaaaa}
\end{array}
\end{eqnarray}
\begin{eqnarray}\label{BD25}
\begin{array}{rcl}
\Delta_{r_{4}^{}}(e_{+})\!\!&=\!\!&e_{+}\otimes\cos\alpha h+\cos\alpha h\otimes e_{+}-
e_{+}'\otimes\sin\alpha h+\sin\alpha h\otimes e_{+}'\;+\phantom{aaaa}
\\[8pt]
&&+\;\displaystyle\frac{\nu\sin\alpha}{2}\Bigl(e_{+}\otimes q_{\imath\alpha}^{h-1}
(e_{+}^{2} +{e'_{+}}^{\!2})+(e_{+}^{2}+ {e'_{+}}^{\!2})q_{\imath\alpha}^{h+1}\otimes
e_{+}\Bigr)\;+
\\[10pt]
&&+\;\displaystyle\frac{\imath\nu\sin\alpha}{2}\Bigl(e_{+}'\otimes
q_{\imath\alpha}^{h-1}(e_{+}^{2}+{e'_{+}}^{\!2})-
(e_{+}^{2}+{e'_{+}}^{\!2})q_{\imath\alpha}^{h+1}\otimes e_{+}'\Bigr)~,
\end{array}
\end{eqnarray}
\begin{eqnarray}\label{BD26}
\begin{array}{rcl}
\Delta_{r_{4}^{}}(e_{+}')\!\!&=\!\!&e_{+}'\otimes\cos\alpha h+\cos\alpha h\otimes e_{+}'+
e_{+}\otimes\sin\alpha h-\sin\alpha h\otimes e_{+}\;+\phantom{aaaa}
\\[8pt]
&&+\;\displaystyle\frac{\nu\sin\alpha}{2}\Bigl(e_{+}'\otimes q_{\imath\alpha}^{h-1}
(e_{+}^{2} +{e'_{+}}^{\!2})+(e_{+}^{2}+ {e'_{+}}^{\!2})\,q_{\imath\alpha}^{h+1}\otimes
e_{+}'\Bigr)\,-
\\[10pt]
&&-\;\displaystyle\frac{\imath\nu\sin\alpha}{2}\Bigl(e_{+}\otimes
q_{\imath\alpha}^{h-1}(e_{+}^{2}+{e'_{+}}^{\!2})-
(e_{+}^{2}+{e'_{+}}^{\!2})\,q_{\imath\alpha}^{h+1}\otimes e_{+}\Bigr)~,
\end{array}
\end{eqnarray}
\begin{eqnarray}\label{BD27}
\begin{array}{rcl}
\Delta_{r_{4}}(e_{-})\!\!&=\!\!& e_{-}\otimes q_{\alpha}^{-h'}+ q_{\alpha}^{h'}\otimes
e_{-}\,+
\\[10pt]
&&+\;\displaystyle\frac{\nu(\sin\alpha)^{-1}}{2}\Bigl(e_{+}q_{\imath\alpha}^{h-\imath
h'}\otimes q_{\alpha}^{-h'}+q_{\alpha}^{h'}\otimes q_{\imath\alpha}^{h+\imath
h'}e_{+}\Bigr)\,+
\\[10pt]
&&+\;\displaystyle\frac{\imath\nu(\sin\alpha)^{-1}}{2}\Bigl(e'_{+}q_{\imath\alpha}^{h-\imath
h'}\otimes q_{\alpha}^{-h'}-q_{\alpha}^{h'}\otimes q_{\imath\alpha}^{h+\imath
h'}e'_{+}\Bigr)\,-
\\[10pt]
&&-\;\displaystyle\frac{\nu(\sin\alpha)^{-1}}{2}\Bigl(e_{+}q_{\alpha}^{h'}\otimes
q_{\imath\alpha}^{h-\imath h'}\Delta_{r_{4}}(q_{\imath\alpha}^{h})+\Delta_{r_{4}}
(q_{\imath\alpha}^{h})\,q_{\imath\alpha}^{h+\imath h'}\otimes
q_{\alpha}^{-h'}e_{+}\Bigr)\,-
\\[10pt]
&&-\;\displaystyle\frac{\imath\nu(\sin\alpha)^{-1}}{2}\Bigl(e'_{+}q_{\alpha}^{h'}\otimes
q_{\imath\alpha}^{h-\imath h'}\Delta_{r_{4}}(q_{\imath\alpha}^{h})-\Delta_{r_{4}}
(q_{\imath\alpha}^{h})\,q_{\imath\alpha}^{h+\imath h'}\otimes
q_{\alpha}^{-h'}e'_{+})\Bigr)~,
\end{array}
\end{eqnarray}
\begin{eqnarray}\label{BD28}
\begin{array}{rcl}
\Delta_{r_{4}}(e'_{-})\!\!&=\!\!&e'_{-}\otimes q_{\alpha}^{-h'}+ q_{\alpha}^{h'}\otimes
e'_{-}\,-
\\[10pt]
&&-\;\displaystyle\frac{\nu(\sin\alpha)^{-1}}{2}\Bigl(e'_{+}q_{\imath\alpha}^{h-\imath
h'}\otimes q_{\alpha}^{-h'}+q_{\alpha}^{h'}\otimes q_{\imath\alpha}^{h+\imath
h'}e'_{+}\Bigr)\,+
\\[10pt]
&&+\;\displaystyle\frac{\imath\nu(\sin\alpha)^{-1}}{2}\Bigl(e_{+}q_{\imath\alpha}^{h-\imath
h'}\otimes q_{\alpha}^{-h'}-q_{\alpha}^{h'}\otimes q_{\imath\alpha}^{h+\imath
h'}e_{+}\Bigr)\,+
\\[10pt]
&&+\;\displaystyle\frac{\nu(\sin\alpha)^{-1}}{2}\Bigl(e'_{+}q_{\alpha}^{h'}\otimes
q_{\imath\alpha}^{h-\imath h'}\Delta_{r_{4}}(q_{\imath\alpha}^{h})+\Delta_{r_{4}}
(q_{\imath\alpha}^{h})\,q_{\imath\alpha}^{h+\imath h'}\otimes
q_{\alpha}^{-h'}e'_{+}\Bigr)\,-
\\[10pt]
&&-\;\displaystyle\frac{\imath\nu(\sin\alpha)^{-1}}{2}\Bigl(e_{+}q_{\alpha}^{h'}\otimes
q_{\imath\alpha}^{h-\imath h'}\Delta_{r_{4}}(q_{\imath\alpha}^{h})-\Delta_{r_{4}}
(q_{\imath\alpha}^{h})\,q_{\imath\alpha}^{h+\imath h'}\otimes
q_{\alpha}^{-h'}e_{+}\Bigr)~.
\end{array}
\end{eqnarray}
Here we remaind that $\xi=\imath\alpha$, $\eta=-2\imath\nu$ where  parameters $\alpha$ and
$\nu$ are real. Explicit formulas of antipodes are given as follows
\begin{eqnarray}\label{BD29}
S_{r_{4}}(q_{\alpha}^{\pm h'})\!\!&=\!\!&q_{\alpha}^{\mp h'}~,
\\[5pt]\label{BD30}
S_{r_{4}}(q_{\imath\alpha}^{\mp
h})\!\!&=\!\!&\Bigl(\bigl(1-2\imath\nu(q_{\imath\alpha}^{2}-1) (e_{+}^2 +
{e'}_{+}^2)\bigr)q_{\imath\alpha}^{h}\Bigr)^{\pm 1}~,
\\[5pt]\label{BD31}
S_{r_{4}}(e_{+}^{})\!\!&=\!\!&-\cos\alpha\,e_{+}^{}+\sin\alpha\,e_{+}'~,
\\[5pt]\label{BD32}
S_{r_{4}}(e_{+}')\!\!&=\!\!&-\cos\alpha\,e_{+}'-\sin\alpha\,e_{+}^{}~,
\end{eqnarray}
\begin{eqnarray}\label{BD33}
\begin{array}{rcl}
S_{r_{4}}(e_{-}^{})\!\!&=\!\!&-\cos\alpha\,e_{-}^{}-\sin\alpha\,e_{-}'\,-
\\[10pt]
&&-\;\displaystyle\frac{\nu(\sin\alpha)^{-1}}{2}\Bigl(e_{+}q_{\imath\alpha}^{h+\imath
h'-1}+q_{\imath\alpha}^{h-\imath h'+1}e_{+}\Bigr)\,+
\\[10pt]
&&+\;\displaystyle\frac{\imath\nu(\sin\alpha)^{-1}}{2}\Bigl(e'_{+}q_{\imath\alpha}^{h+\imath
h'-1}-q_{\imath\alpha}^{h-\imath h'+1}e'_{+}\Bigr)\,+
\\[10pt]
&&+\;\displaystyle\frac{\nu(\sin\alpha)^{-1}}{2}\Bigl(e_{+}q_{\imath\alpha}^{-\imath
h'-1}S_{r_{4}}(q_{\imath\alpha}^{h})+S_{r_{4}}(q_{\imath\alpha}^{h})
q_{\imath\alpha}^{\imath h'+1}e_{+}\Bigr)\,-
\\[10pt]
&&-\;\displaystyle\frac{\imath\nu(\sin\alpha)^{-1}}{2}\Bigl(e'_{+}q_{\imath\alpha}^{-h'-1}
S_{r_{4}}(q_{\imath\alpha}^{h})- S_{r_{4}}(q_{\imath\alpha}^{h})q_{\imath\alpha}^{\imath
h'+1}e'_{+}\Bigr)~,
\end{array}
\end{eqnarray}
\begin{eqnarray}\label{BD34}
\begin{array}{rcl}
S_{r_{4}}(e_{-}^{})\!\!&=\!\!&-\cos\alpha\,e_{-}'+\sin\alpha\,e_{-}\,+
\\[10pt]
&&+\;\displaystyle\frac{\nu(\sin\alpha)^{-1}}{2}\Bigl(e'_{+}q_{\imath\alpha}^{h+\imath
h'-1}+q_{\imath\alpha}^{h-\imath h'+1}e'_{+}\Bigr)\,+
\\[10pt]
&&+\;\displaystyle\frac{\imath\nu(\sin\alpha)^{-1}}{2}\Bigl(e_{+}q_{\imath\alpha}^{h+\imath
h'-1}-q_{\imath\alpha}^{h-\imath h'+1}e_{+}\Bigr)\,-
\\[10pt]
&&-\;\displaystyle\frac{\nu(\sin\alpha)^{-1}}{2}\Bigl(e_{+}'q_{\imath\alpha}^{-\imath
h'-1}S_{r_{4}}(q_{\imath\alpha}^{h})+S_{r_{4}}(q_{\imath\alpha}^{h})
q_{\imath\alpha}^{\imath h'+1}e_{+}'\Bigr)\,-
\\[10pt]
&&-\;\displaystyle\frac{\imath\nu(\sin\alpha)^{-1}}{2}\Bigl(e_{+}^{}
q_{\imath\alpha}^{-\imath h'-1} S_{r_{4}}(q_{\imath\alpha}^{h})-
S_{r_{4}}(q_{\imath\alpha}^{h})q_{\imath\alpha}^{\imath h'+1}e_{+}^{}\Bigr)~.
\end{array}
\end{eqnarray}

\setcounter{equation}{0}
\section{Outlook}

In this paper we presented in detail the Hopf algebra structures describing  three
quantum deformations of $D=4$ Lorentz algebra $\mathfrak{o}(3,1)$. Two of them are
obtained by twisting of the standard $q$-deformation $U_{q}(\mathfrak{o}(3,1))$. For the
first twisted $q$-deformation an Abelian twist depending on Cartan generators of
$\mathfrak{o}(3,1)$ was used. The second example of twisting provides a quantum
deformation of Cremmer-Gervais type for the Lorentz algebra. By twist quantization
techniques we obtained explicit formulae for the deformed coproducts and antipodes of the
$\mathfrak{o}(3,1)$-generators. For the sake of completeness  we also incorporated here
non-standard Jordanian deformation, while the remaining extended Jordanian twist has been
considered with details in \cite{BLT1, BLT2}.

The next step in the programme of explicit description of quantum deformations of the
relativistic symmetries is to look for twists of quantum deformations of the Poincar\'{e}
algebra, described by modified classical $r$-matrices. For that purpose one should find
firstly full classification of $D=4$ Poincar\'{e} modifed classical $r$-matrices by
completing the results in \cite{Zak2}. It should be add that recently (see
\cite{BoLuTo3}) we also started the systematic study of twists describing quantum
deformations of Poinca\'{r}e superalgebra.

\setcounter{equation}{0}
\section{Appendix 
}

Here we consider some specialization of a $q$-deformed Hadamard lemma \footnote{Hadamard
lemma is sometimes called Baker-Campbell-Hausdorff formula.}, which is main tool for
calculations of our results in Sect.5.

Let $A$ and $B$ be two arbitrary elements of some quantum algebra and let $\exp_{q}(A)$
be a formal $q$-exponential (\ref{ct16}) of the element $A$. The formal $q$-exponential
$\exp_{q^{-1}}(-A)$ is inverse to $\exp_{q}(A)$, i.e. $\bigl(\exp_{q}(A)\bigr)^{-1}=
\exp_{q^{-1}}(-A)$. The $q$-analog of  Hadamard formula is given as follows (see
\cite{KhTo1})
\begin{eqnarray}\label{A1}
\begin{array}{rcl}
\exp_{q}(A)\,B\bigl(\exp_{q}(A)\bigr)^{-1}\!\!&=\!\!&\exp_{q}(A)\,B\exp_{q^{-1}}(-A)\;
\equiv\;\bigl(\mathop{\rm Ad}\exp_{q}(A)\bigr)(B)\;=\;
\\[10pt]
\!\!&=\!\!&\displaystyle\Bigl(\sum_{n\geq 0}\frac{1}{(n)_{q}!}(\mathop{{\rm
ad}_q}A)^n\Bigr)(B)\;=\;\bigl(\exp_{q}(\mathop{{\rm ad}_q}A)\bigr)(B)~,
\end{array}
\end{eqnarray}
where one sets
\begin{eqnarray}\label{A2}
\begin{array}{rcl}
(\mathop{{\rm ad}_q}A)^0(B)\!\!&\equiv\!\!&B~,\quad(\mathop{{\rm ad}_q}(A))^1(B)\;\equiv
\;[A,\,B]~,\quad(\mathop{{\rm ad}_q}(A))^2(B)\;\equiv\;[A,\,[A,\,B]]_{q}~,
\\[10pt]
(\mathop{{\rm ad}_q}(A))^3(B)\!\!&\equiv\!\!&[A,\,[A,\,[A,\,B]]_{q}]_{q^2}~,\ldots,
(\mathop{{\rm ad}_q}(A))^{n+1}(B)\;=\;[A,\,(\mathop{{\rm ad}_q}(A))^{n}(B)]_{q^n}~.
\end{array}
\end{eqnarray}
Here the $q$-brackets $[\cdot,\cdot]_{q'}$ means $[C,\,D]_{q'}\;=\;CD-q'DC$~.

The point is to obtain  results in a compact (finite) form. It can be achieved here due
to the following facts:
\begin{itemize}
\item {\em Assume additionally  that $[A, B]_q=0$. Then}
\begin{eqnarray}\label{A3}
\exp_{q}(A)\,B\bigl(\exp_{q}(A)\bigr)^{-1}\!\!&=\!\!&\tilde{X}B\;=\;BX~.
\end{eqnarray}
\item {\em Assume  $[A, B]_{q^{-1}}=0$.}
Thus
\begin{eqnarray}\label{A4}
\exp_{q}(A)\,B\bigl(\exp_{q}(A)\bigr)^{-1}\!\!&=\!\!&X^{-1}B\;=\;B\tilde{X}^{-1}~.
\end{eqnarray}
{\em where $X=1+(q-1)A$,  $\tilde X=1-(q^{-1}-1)A$}~.
\end{itemize}
In our case we have to substitute $A\rightarrow\mathbb{A}=\eta
E_{1+}q_{\xi}^{H_{1}+H_{2}}\otimes q_{\xi}^{H_{1}+H_{2}}E_{2+}$ or $A=\eta E_{1+}E_{2+}$
(cf. (\ref{BD15}) or (\ref{BD22})). This, due to unitarity, implies $\tilde X= X^*$.
Therefore, each formula (\ref{BD8})--(\ref{BD21}) admits its counterpart. For example,
left handed version of (\ref{BD10}) reads
\begin{equation}\label{A5}
\Delta_{r_{4}}(q_{\xi}^{-H_{1}-H_2})\;=\;\mathbb{X}^*\,q_{\xi}^{-H_{1}-H_2}\otimes
q_{\xi}^{-H_{1}-H_2}~.
\end{equation}

Further we need
\begin{eqnarray}\label{A6}
[q_{\xi}^{aH_{1}+bH_{2}}\otimes q_{\xi}^{cH_{1}+dH_{2}},\,\mathbb{A}]_{q_{\xi}^{a+d}}
\!\!&=\!\!&0\quad\forall\,a,b,c,d\in\mathbb{C}~.
\end{eqnarray}
Analogical  expression
\begin{equation}\label{A7}
[q_{\xi}^{aH_{1}+dH_{2}},\,A]_{q_{\xi}^{a+d}}\;=\;0
 \end{equation}
is valid for $A=\eta E_{1+}E_{2+}$. Now specialization to three cases $a+d=0,\, 2\;
\mbox{or} -2$ together with (\ref{A3}) and (\ref{A4}) gives rise to formulae
(\ref{BD8})--(\ref{BD14}) and (\ref{BD17})--(\ref{BD21}).

In order to calculate (\ref{BD14}), for example, one decomposes
\begin{equation}\label{A8}
q_{\xi}^{H_{2}-H_{1}}\otimes E_{2-}\;=\;\big( q_{\xi}^{-2H_{1}-2H_{2}}\otimes E_{2-}
q_{\xi}^{H_{1}+H_{2}}\big) \big( q_{\xi}^{H_{1}+3H_{2}}\otimes
q_{\xi}^{-H_{1}-H_{2}}\big) \end{equation} The second term commutes with the twist while
the first one can be treated by (\ref{A4}).

Another interesting property can be realized by applying an operator $m\circ (id\otimes
S_{r'_{4}}^{})$ to (\ref{BD10}). Then comparison  with (\ref{BD18}) gives
\begin{equation}\label{A9}
m\circ (id\otimes S_{r'_{4}}^{})(\mathbb{X})\;=\;X^{-1}~.
\end{equation}
\subsection*{Acknowledgments}
The paper has been supported by MNiSW grant NN202 318534 (A.B., J.L., V.N.T.)  and the
grant RFBR-08-01-00392 (V.N.T.), and the French National Research Agency grant
NT05-241455GIPM (V.N.T.). The first author acknowledges financial support from the
Bogoliubov-Infeld Program 2008. The third author would like to thank Institute for
Theoretical Physics, University of Wroc{\l}aw for hospitality.

\end{document}